\documentstyle[psfig]{mn}
 
\newif\ifAMStwofonts 
\ifoldfss 
  \ifCUPmtlplainloaded \else 
    \NewTextAlphabet{textbfit} {cmbxti10} {} 
    \NewTextAlphabet{textbfss} {cmssbx10} {} 
    \NewMathAlphabet{mathbfit} {cmbxti10} {} % for math mode 
    \NewMathAlphabet{mathbfss} {cmssbx10} {} %  "   "    " 
  \fi 
  \ifAMStwofonts 
    \ifCUPmtlplainloaded \else 
      \NewSymbolFont{upmath} {eurm10} 
      \NewSymbolFont{AMSa} {msam10} 
      \NewMathSymbol{\upi}     {0}{upmath}{19} 
      \NewMathSymbol{\umu}     {0}{upmath}{16} 
      \NewMathSymbol{\upartial}{0}{upmath}{40} 
      \NewMathSymbol{\leqslant}{3}{AMSa}{36} 
      \NewMathSymbol{\geqslant}{3}{AMSa}{3E}

       \let\le=\leqslant 
        
    \fi 
  \fi 
\fi % End of OFSS 
 
\ifnfssone 
  \newmathalphabet{\mathit} 
  \addtoversion{normal}{\mathit}{cmr}{m}{it} 
  \addtoversion{bold}{\mathit}{cmr}{bx}{it} 
  \newmathalphabet{\mathbfit} % math mode version of \textbfit{..} 
  \addtoversion{normal}{\mathbfit}{cmr}{bx}{it} 
  \addtoversion{bold}{\mathbfit}{cmr}{bx}{it} 
  \newmathalphabet{\mathbfss} % math mode version of \textbfss{..} 
  \addtoversion{normal}{\mathbfss}{cmss}{bx}{n} 
  \addtoversion{bold}{\mathbfss}{cmss}{bx}{n} 
  \ifAMStwofonts 
    \ifCUPmtlplainloaded \else 
      \UseAMStwoboldmath 
      \makeatletter 
      \new@mathgroup\upmath@group 
      \define@mathgroup\mv@normal\upmath@group{eur}{m}{n} 
      \define@mathgroup\mv@bold\upmath@group{eur}{b}{n} 
      \edef\UPM{\hexnumber\upmath@group} 
      \new@mathgroup\amsa@group 
      \define@mathgroup\mv@normal\amsa@group{msa}{m}{n} 
      \define@mathgroup\mv@bold\amsa@group{msa}{m}{n} 
      \edef\AMSa{\hexnumber\amsa@group} 
      \makeatother 
      \mathchardef\upi="0\UPM19 
      \mathchardef\umu="0\UPM16 
      \mathchardef\upartial="0\UPM40 
      \mathchardef\leqslant="3\AMSa36 
      \mathchardef\geqslant="3\AMSa3E 

       \let\le=\leqslant 

    \fi 
  \fi 
\fi % End of NFSS release 1 
 
\ifnfsstwo 
  \DeclareMathAlphabet{\mathbfit}{OT1}{cmr}{bx}{it} 
  \SetMathAlphabet\mathbfit{bold}{OT1}{cmr}{bx}{it} 
  \DeclareMathAlphabet{\mathbfss}{OT1}{cmss}{bx}{n} 
  \SetMathAlphabet\mathbfss{bold}{OT1}{cmss}{bx}{n} 
  \ifAMStwofonts 
    \ifCUPmtlplainloaded \else 
      \DeclareSymbolFont{UPM}{U}{eur}{m}{n} 
      \SetSymbolFont{UPM}{bold}{U}{eur}{b}{n} 
      \DeclareSymbolFont{AMSa}{U}{msa}{m}{n} 
      \DeclareMathSymbol{\upi}{0}{UPM}{"19} 
      \DeclareMathSymbol{\umu}{0}{UPM}{"16} 
      \DeclareMathSymbol{\upartial}{0}{UPM}{"40} 
      \DeclareMathSymbol{\leqslant}{3}{AMSa}{"36} 
      \DeclareMathSymbol{\geqslant}{3}{AMSa}{"3E} 

       \let\le=\leqslant 

    \fi 
  \fi 
\fi % End of NFSS release 2 
 
\ifCUPmtlplainloaded \else 
  \ifAMStwofonts \else % If no AMS fonts 
    \def\upi{\pi} 
    \def\umu{\mu} 
    \def\upartial{\partial} 
  \fi 
\fi

\title{On the reliability of the semiempirical RR Lyrae Period - Blue Amplitude 
- V-band Luminosity relation.} 
\author[R. De Santis and S. Cassisi] 
       {R. De Santis$^1$ and S. Cassisi $^{1,2}$ 
\\ 
        $^1$ INAF - Osservatorio Astronomico di Teramo, via M. Maggini, 64100 Teramo,  
             Italy~~cassisi@te.astro.it 
\\ 
        $^2$ Max-Planck fur Astrophysik, Karl-Schwarzschild-Strasse 1, 85741 Garching, Germany 
	     } 
\date{} 
 
\begin{document} 
 
\maketitle 
 
\label{firstpage} 
 
\begin{abstract} 
 
We investigate the accuracy and reliability of the semiempirical period - blue amplitude -  
V-band luminosity relationship for ab-type RR Lyrae stars originally obtained by  
Castellani \& De Santis (1994) and De Santis (1996). We infer that the
zero point of this relationship does depend on the metallicity, by
studying a sample of both 
field and cluster variables.  
We also show that the use of this relationship can still be useful for those stellar  
systems showing an intrinsic metallicity spread, since in this case the metallicity effect 
has a negligible effect on the final distance modulus estimate. 
 
We compare the adopted semiempirical relationship with the fully empirical one recently  
provided by Kov\'acs \& Walker (2001). When the zero point of the latter  
relation is fixed consistently with the former one, the two equations are equivalent. 
 
By appling the semiempirical period - blue amplitude - V-band luminosity relation, as well as the technique 
proposed by Cassisi, De Santis \& Piersimoni (2001), to the globular cluster $\omega$~Cen, we show 
that the empirical slope of the relationship between the mass of the fundamental RR Lyrae pulsators and their  
metallicity, is in fair agreement with the one predicted by updated evolutionary models for 
Horizontal-Branch stars. 
 
\end{abstract} 
 

\section{Introduction} 
 
The traditional distance indicator for Population II stellar systems 
is the magnitude of RR Lyrae variables; for this reason, several  
observational and theoretical investigations have been devoted to this class of variables 
(Bono, Castellani \& Marconi 2000, Clement et al. 2001). 

In spite of the large body of work devoted to their
study, some relevant questions are still unanswered.
One of the most important problems is the lack 
of a general agreement about both the slope and the zero-point 
of the absolute magnitude - metallicity ($M_V(RR) - [Fe/H]$)
relationship, characteristic of the RR Lyrae stars
(e.g., Caputo 1997, Gratton 1998, Cassisi, De Santis\footnote{We inform with great sorrow 
that R. De Santis died on the 4th March 2002} \& Piersimoni 2001, Benedict et 
al. 2002 and references therein). 
Some methods, like Baade-Wesselink and statistical parallax analyses 
applied to field RR Lyrae stars, plus observations of field Horizontal 
Branch (HB) stars with parallax measurements (but with large
parallax errors --  Gratton 1998) support the {\sl short} 
distance scale. On the other hand, the pulsational properties of cluster 
RR Lyrae stars (Sandage 1993), the main-sequence fitting to local subdwarfs 
(Gratton et al. 1997), the calibration of HB luminosity obtained  
by using the Cepheid distance modulus of the Large Magellanic Cloud 
(Walker 1992), and analysis based on double-mode RR Lyrae (Kov\'acs 2000, Popielski, 
Dziembowski \& Cassisi 2000) support the {\sl long} scale. 
 
One possible explanation of such a disagreement could be the existence 
of a true luminosity difference between field and cluster HB stars, as suggested 
by Gratton (1998) on the basis of an Hipparcos calibration of the absolute 
magnitude of field HB stars. However, this evidence is not supported 
by the analyses performed by Catelan (1998), De Santis \& Cassisi (1999, hereinafter DC) 
and Carretta, Gratton \& Clementini (2000). The origin of the distance dichotomy is  
therefore still unexplained and it has a big impact 
on a wide range of astrophysical problems such as globular clusters (GCs) age determinations 
and the extragalactic distances measurements.  
 
A significant contribution to the solution of this problem can be provided by the analysis of the  
pulsational properties of RR Lyrae stars both in the Galaxy and in the Large Magellanic Clouds.  
A first step towards this direction was made by Sandage, Katem \& Sandage (1981) who suggested  
the existence of a tight correlation between temperature and amplitude of RR Lyrae stars.  
More recently, Caputo \& De Santis (1992) showed the existence of a  
clear correlation between period, blue amplitude and light-mass ratio of the 
variables. The reliability of these and similar relationships is important, since 
period and amplitude can be measured with high accuracy, regardless of uncertainties 
on both distance modulus and reddening. In this field, a pivotal importance is played 
by any relation connecting pulsational properties to the intrinsic luminosity 
of RR Lyrae stars. 
 
Theoretical support for the existence of a correlation between visual magnitude, period and blue 
amplitude (hereinafter {\it PLA} relation) for fundamental RR Lyrae
($RR_{ab}$) pulsators has been presented  
by Castellani \& De Santis (1994) and by De Santis (1996). They provided also a semiempirical 
calibration of this relationship by adopting theoretical pulsational models and  
the observational database available at that time. However, until now, we lack a definitive
assessment of the reliability of such relationship. 
 
This paper is the third of a series investigating how the pulsational  
properties of RR Lyrae stars can be used to constrain their intrinsic luminosity. 
In particular, DC have used the pulsational behaviour of $RR_{ab}$ stars to obtain  
an accurate estimate of the absolute bolometric luminosity of ZAHB stars in GCs. 
It is worth noticing that their results do not depend 
on the underlying evolutionary models of HB stars, 
and this occurrence allowed them to perform a significant comparison 
with recent theoretical evaluations of the ZAHB luminosity. 
This, in turn, is important in order to properly evaluate the reliability 
of current theoretical models of low-mass, He-burning stars (see also 
the discussion in Vandenberg et al. 2000) . 
 
Cassisi, De Santis \& Piersimoni (2001, hereinafter CDP) have adopted the same method 
outlined by DC, in order to derive the absolute visual magnitude of the ZAHB 
within the RR Lyrae instability strip, for a
sample of galactic GCs with accurate photometric 
data for both variable and non-variable HB stars.  
After applying a correction for the difference between the mean RR Lyrae magnitude  
and the ZAHB one, they derived a  $<M_V(RR)> - [Fe/H]$ relation and compared it 
with the most recent empirical ones (see also Caputo et al. 1999).  
They also discussed a method for determining the GCs distance based only on the pulsational properties  
of their RR Lyrae population. The advantage of this method is that it does not need an estimate 
of the ZAHB level, which is particularly difficult task in the case of 
GCs with blue HB, whose RR Lyrae are suspected to be evolved stars.  
 
The approaches developed by DC and CDP 
require a preliminary estimate of the cluster metallicity in order to determine
the appropriate mass range for $RR_{ab}$ stars. Due to the non-negligible 
uncertainties affecting both the GC metallicity scale (see, i.e., Rutledge et al. 1997 and 
Vandenberg et al. 2000) and their $\alpha$-elements distribution, this reduces 
the accuracy of the method. Therefore, in the present work we investigate the possibility 
to estimate the absolute visual magnitude of $RR_{ab}$ stars 
by adopting a magnitude-period-amplitude relation. This would not require 
a preliminary evaluation of the stellar metallicity. 
 
In the next section we briefly review the semiempirical PLA relationship adopted in present  
analysis. In section 3, we apply this relationship to a sample of field RR Lyrae stars,  
in order to investigate on its accuracy when applied to single stars.  
In section 4, we follow the same approach in case of a selected 
sample of galactic GCs, and the derived distance moduli are compared with those provided 
by CDP. In section 5, the semiempirical PLA relation is compared with the 
empirical one by Kovacs \& Walker (2001). An application of the method to  
stellar systems showing a spread in the metallicity, like the  
GC $\omega$~Cen, is shown in section 6. A brief discussion and conclusions  
follow in the last section. 
 
\section{\bf The semiempirical PLA relation.} 
 
Castellani \& De Santis (1994) showed the existence of a tight relation 
between effective temperature, period and blue amplitude. By combining this relation 
$T_e=T_e(P,A_B)$ with the fundamental pulsational equation correlating the period of the  
variable with its main evolutionary properties such as mass, luminosity and 
effective temperature (van Albada \& Baker 1971), they derived 
a relationship between the RR Lyrae intrinsic luminosity, period, blue amplitude, mass 
and visual bolometric correction: 
 
\smallskip 
$$M_V=-1.842\cdot\log{P}-0.137\cdot{A_B}+~~~~~~~~~~~~~~~~~~~~$$ 
$$~~~~~~~~~~~~~~~~~~~~~-(2.02\cdot\log{M/M_\odot}+BC_V+0.19)\eqno(1)$$  
\smallskip 
\noindent 
 
Making use of theoretical results available at that time for both the 
mass of fundamental pulsators and bolometric corrections, they estimated 
the term in parentheses to be about 0.2 mag. They derived also 
an empirical calibration of the zero point of equation (1), by
applying it to a sample 
of RR Lyrae stars with known P, $A_B$ and absolute visual magnitude, keeping as a free parameter 
the additional term.  
For this calibration, the results of Baade-Wesselink analyses summarized by Sandage \& Cacciari (1990) 
were used. The final zero point was equal to 
0.45 mag, quite different from the one obtained on the basis of
stellar evolution
results. Nevertheless, an important result was the evidence that the zero point 
of relation (1) does not seem to depend on the star metallicity. 
This topic was further addressed by De Santis (1996, hereinafter DS),
who analyzed data for 27 $RR_{ab}$ in the cluster $\omega$~Cen (Sandage 1990).  
DS showed that the dependence on the metallicity of the term in parentheses in
equation (1) was completely negligible. 
 
The reliability of this latter result relies on the accuracy of  
the adopted metallicity scale (Sandage 1990). In the same work, by using current  
theoretical results for the mass of RR Lyrae variables, the zero point  
was fixed to 0.31 mag, obtaining: 
 
$$M_V=-1.842\cdot\log{P}-0.137\cdot{A_B}+0.31\eqno(2)$$ 
 
\noindent 
Since in the following we want to compare the GCs distance moduli
obtained by using the PLA relation with the ones obtained by CDP which made use of a more 
updated evolutionary and pulsational scenario,  
we have rederived equation (2) by using 
the updated fundamental pulsational equation provided by Bono et al. (1997).  
The final semiempirical relation is then: 
 
$$M_V=-1.842\cdot\log{P}-0.137\cdot{A_B}+0.26\eqno(3)$$  
 
\noindent 
The reliability of this relation relies on the accuracy 
of the empirical temperature scale for RR Lyrae stars derived by DS, and adopted by DC and CDP.  
This issue has been recently investigated by De Santis (2001). 
 
Due to the existence of significant uncertainties between different 
metallicity scales (Zinn \& West 1984, Carretta \& Gratton 1997, Rutledge, Stetson \& Hesser 1997)  
as well as the availability of new metallicity estimates 
for a large RR Lyrae sample, it is worth investigating the dependence 
of the zero point of equation (3) on metallicity. We address this topic in the 
next section. 
 
%%%%%%%%%%%%%%%%%%%%%%%%%%%%%%%%%%%%%%%%%%%%%%%%%%%%%%%%%%%%%%%%%%% 
%%%%                         FIGURE 1                           %%% 
%%%%%%%%%%%%%%%%%%%%%%%%%%%%%%%%%%%%%%%%%%%%%%%%%%%%%%%%%%%%%%%%%%% 
\begin{figure} 
\psfig{figure=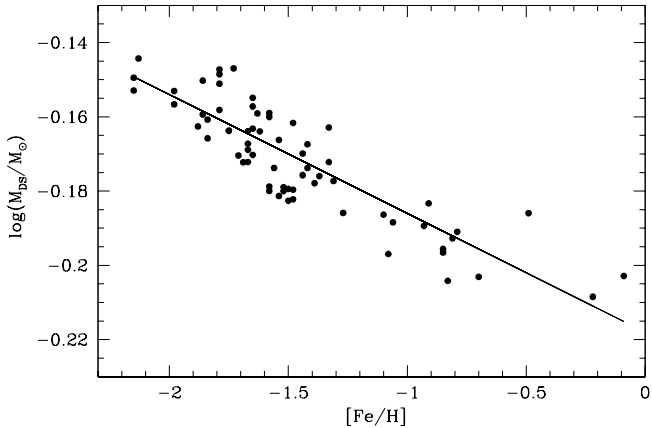,height=9cm,width=9cm} 
\caption{The behaviour of the mass of 
the field variable stars in the Lub's sample - obtained by using the 
absolute visual magnitude estimates provided by equation 3), as a function 
of their metallicity. The solid line corresponds to a linear regression  
to the data (see text for more details).} 
\end{figure} 
 
\section{\bf A check of the accuracy of the PLA relation.} 
 
When the terms in parentheses in equation (1) are fixed, 
the use of the PLA relation does not require an estimate of the pulsator mass.  
On the contrary, the approach adopted by DC and  
CDP for estimating the intrinsic bolometric luminosity of the ZAHB is
based on {\sl a priori}  
estimates of the allowed mass range for fundamental pulsators. 
In the above papers, it has been shown that a change of about 
$+0.01M_\odot$ for both the upper and lower limit of the mass of
the fundamental pulsators 
implies a change in the value of the cluster distance modulus of the order of $+0.01$ mag. 
 
One can therefore check the accuracy of the PLA relation by comparing the 
pulsator mass obtained by adopting the absolute visual magnitude estimate from the PLA relation,  
with the value provided by evolutionary HB models. 
One can estimate the intrinsic bolometric luminosity of a star  
by using the relation:  
 
$$\log{L/L_\odot}=-0.4\cdot(M_V-M_{V,\odot}+BC_V-BC_{V,\odot})\eqno(4)$$ 
 
\noindent 
For each variable in the sample the quantity $(BC_V-BC_{V,\odot})$ can be  
derived from equations (3) listed by DC and CDP, while its effective temperature 
can be obtained by using the temperature scale by DS. 
Once obtained the absolute visual magnitude by using relation (3) 
and estimated its $T_{eff}$ value, it is easy to derive the mass of the variable  
from the fundamental pulsational equation (Bono et al. 1997). In the following  
we adopt this approach for both field and cluster RR Lyrae. 
 
\subsection{\bf Field RR Lyrae variables.} 
 
By applying this method to a sample of field $RR_{ab}$ stars (Lub 1977), we have derived  
an estimate of the mass for each variable. These values are plotted in figure 1 as a function 
of [Fe/H]. From these data the slope of $\log{M}_{DS}$  
(hereinafter $M_{DS}$ denotes the mass 
of the variable when the absolute visual magnitude has been derived according to relation 3) 
versus the metallicity is equal to $-0.030\pm0.001$. This slope is
half of the value predicted by current HB models (see DC).  
 
For a metallicity $[Fe/H]\approx-2.0$ theoretical models suggest a mass range for 
fundamental pulsators $0.70\le{M_{RR}}/M_\odot\le0.80$, whereas from  
relation 3) we obtain $M_{DS}\approx0.72M_\odot$; when $[Fe/H]\approx-1.5$ 
evolutionary theory suggests $0.66\le{M_{RR}}/M_\odot\le0.72$, whereas relation 3) predicts 
$M_{DS}\approx0.69M_\odot$; in the metal-rich regime $[Fe/H]\approx-0.8$ we can derive 
from HB models $0.60\le{M_{RR}}/M_\odot\le0.63$ which has to be compared with the value 
$M_{DS}\approx 0.65M_\odot$. A metallicity range of $\pm0.1$ dex was adopted in calculating the  
above values of $M_{DS}$. 
 
By adopting the mass range for $RR_{ab}$ stars provided by updated evolutionary predictions, 
comparing the quoted values for $M_{DS}$ with these mass ranges and remembering  
that $\Delta{M_{RR}}=0.01M_\odot$ implies an uncertainty of $\Delta{M_V}=0.01$ mag,  
one can estimate that the {\it maximum} systematic error in the
estimated absolute V magnitude  
is $+0.02/-0.08$ mag in the metal-poor regime, $+0.03/-0.03$ mag for intermediate metallicity variables,  
and $+0.05$ mag for metal-rich stars.  
 
Thus, even if the metallicity is unknown, the maximum systematic error affecting the absolute V magnitude 
provided by the PLA relation is $+0.05/-0.08$ mag. 
 
\subsection{\bf RR Lyrae variables in galactic GCs.} 
 
CDP have shown that by adopting the pulsational properties of GC RR Lyrae stars,  
it is possible to estimate the GC distances with high accuracy ($\Delta(m-M)_V\le0.05$ mag).  
In this section, we compare results obtained  in that work with 
the ones we obtain by using the PLA relation. 
 
For each cluster in the CDP\footnote{In the present work we have added to the  
GCs sample adopted by CDP two additional clusters, namely, M54 (Layden \& Sarajedini 2000) and 
NGC6934 (Kaluzny et al. 2001).} sample, the {\sl mean} mass of the variables ($M_{CDP}$)  
has been obtained by using, for each variable, the approach previously outlined and  
the $M_V$ value estimated from the CDP distance modulus. 
 
These values are listed in column 4 of table 1. As one can expect (see the CDP approach), the mean mass  
results to be equal to the central value of the allowed mass range for fundamental pulsators  
provided by evolutionary HB models. 
 
%%%%%%%%%%%%%%%%%%%%%%%%%%%%%%%%%%%%%%%%%%%%%%%%%%%%%%%%%%%%%%%%%%% 
%%%%                         FIGURE 2                           %%% 
%%%%%%%%%%%%%%%%%%%%%%%%%%%%%%%%%%%%%%%%%%%%%%%%%%%%%%%%%%%%%%%%%%% 
\begin{figure} 
\psfig{figure=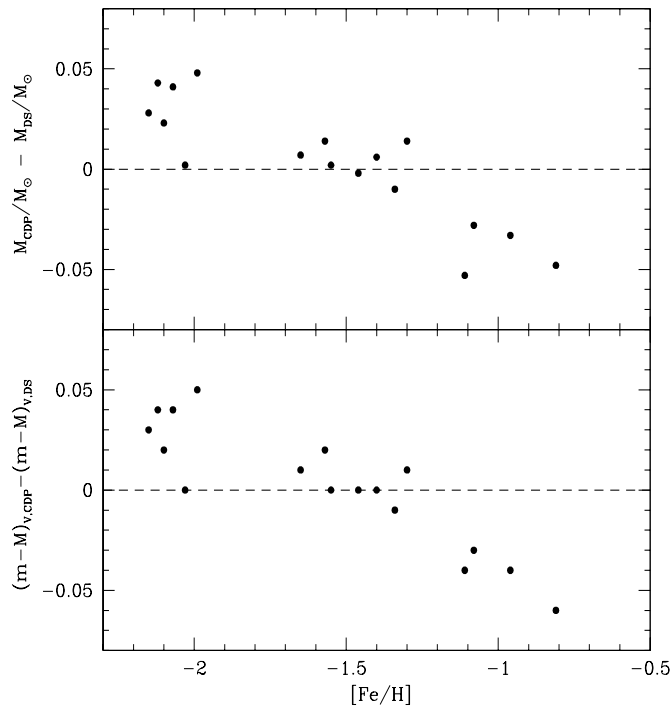,height=10cm,width=9cm} 
\caption{{\sl Upper panel}:  The behaviour with the metallicity of the difference 
between the mean mass value for the clusters' RR Lyrae stars in the CDP sample 
as obtained by estimating their absolute visual magnitude from the distance modulus obtained by CDP  
($M_{CDP}$), and by using the PLA relation ($M_{DS}$).  
{\sl Bottom panel}: As upper panel but for the difference between the distance modulus by CDP and 
the one obtained by using the PLA relation (equation 3).} 
\end{figure} 
%%%%%%%%%%%%%%%%%%%%%%%%%%%%%%%%%%%%%%%%%%%%%%%%%%%%%%%%%%%%%%%%%%% 
 
%%%%%%%%%%%%%%%%%%%%%%%%%%%%%%%%%%%%%%%%%%%%%%%%%%%%%%%%%%%%%%%%%%%%%%%%%%%%%%%%%%% 
%%%%%%%                            TABLE 1                                        % 
%%%%%%%%%%%%%%%%%%%%%%%%%%%%%%%%%%%%%%%%%%%%%%%%%%%%%%%%%%%%%%%%%%%%%%%%%%%%%%%%%%% 
\begin{table*} 
%\centering 
\caption[]{The main properties of the RR Lyrae population in the selected sample of 
globular clusters.\label{tab1}} 
\begin{tabular}{lrccccc} 
NGC   & Name  & [Fe/H] &  $M_{CDP}/M_\odot$  & $M_{DS}/M_\odot$ & 
$(m-M)_{V}^{CDP}$ & $(m-M)_V^{DS}$ \\ 
 
6171  & M107  &  -0.87 & 0.607 & 0.655 &  14.98$\pm$0.02 & 15.04$\pm$0.07\\ 
6362  &       &  -0.96 & 0.625 & 0.658 &  14.64$\pm$0.01 & 14.68$\pm$0.04\\ 
1851  &       &  -1.08 & 0.642 & 0.670 &  15.51$\pm$0.01 & 15.54$\pm$0.05\\ 
5904  & M5    &  -1.11 & 0.636 & 0.665 &  14.47$\pm$0.03 & 14.46$\pm$0.03\\ 
6981  & M72   &  -1.30 & 0.689 & 0.675 &  16.42$\pm$0.01 & 16.41$\pm$0.04\\ 
5272  & M3    &  -1.34 & 0.664 & 0.674 &  15.10$\pm$0.01 & 15.11$\pm$0.02\\ 
6934  &       &  -1.40 & 0.684 & 0.678 &  16.31$\pm$0.05 & 16.31$\pm$0.04\\ 
      & IC4499&  -1.46 & 0.682 & 0.684 &  17.10$\pm$0.05 & 17.10$\pm$0.02\\ 
6715  & M54   &  -1.55 & 0.691 & 0.689 &  17.62$\pm$0.05 & 17.62$\pm$0.03\\ 
6333  & M9    &  -1.57 & 0.698 & 0.684 &  15.75$\pm$0.05 & 15.73$\pm$0.05\\ 
6809  & M55   &  -1.65 & 0.702 & 0.695 &  13.90$\pm$0.05 & 13.89$\pm$0.07\\ 
4590  & M68   &  -1.99 & 0.756 & 0.708 &  15.19$\pm$0.03 & 15.14$\pm$0.04\\ 
5466  &       &  -2.03 & 0.713 & 0.711 &  16.05$\pm$0.03 & 16.05$\pm$0.03\\ 
6426  &       &  -2.07 & 0.760 & 0.719 &  17.80$\pm$0.07 & 17.76$\pm$0.06\\ 
5053  &       &  -2.10 & 0.744 & 0.721 &  16.13$\pm$0.05 & 16.11$\pm$0.07\\ 
7078  & M15   &  -2.12 & 0.754 & 0.711 &  15.36$\pm$0.03 & 15.32$\pm$0.04\\ 
6341  & M92   &  -2.15 & 0.743 & 0.715 &  14.71$\pm$0.05 & 14.68$\pm$0.05\\ 
\end{tabular} 
\end{table*} 
%%%%%%%%%%%%%%%%%%%%%%%%%%%%%%%%%%%%%%%%%%%%%%%%%%%%%%%%%%%%%%%%%%%%%%%%%%%%%%%%%% 
 
By following the same approach outlined in previous section, we have estimated the mean mass ($M_{DS}$)  
of the variables in each cluster. These values are listed in column 5 
of Table 1) and their differences with respect to the $M_{CDP}$ values are plotted in figure 2  
as a function of the metallicity.  
We obtain that for metal-poor clusters $M_{CDP}$ values are larger than 
$M_{DS}$ by about $0.03-0.04M_\odot$, whereas for metal-rich clusters $M_{CDP}$ values are smaller 
than $M_{DS}$ by about the same amount. It is also worth noticing that for 
intermediate-metallicity clusters ($[Fe/H]\approx-1.3$), the two mass estimates  
are in fine agreement.  
 
As expected, the same trend is found for the distance moduli obtained by 
using alternatively the CDP approach ($(m-M)_V^{CDP}$) or the PLA relation ($(m-M)_V^{DS}$) (see table 1).  
This is shown in the bottom panel of figure 2:  
for metal-poor clusters $(m-M)_V^{DS}$  are lower than $(m-M)_V^{CDP}$ by about 
0.04 mag, whereas for metal-rich clusters $(m-M)_V^{DS}$  are larger than $(m-M)_V^{CDP}$  
by the same amount; for intermediate-metallicity clusters the agreement between 
the two distance estimates is fine. 
We conclude that, in the metallicity range $-2.2\le[Fe/H]\le-0.8$ the  
distance moduli obtained by using equation 3) are in agreement with the CDP  
estimates within $\pm0.04$ mag. 
 
%%%%%%%%%%%%%%%%%%%%%%%%%%%%%%%%%%%%%%%%%%%%%%%%%%%%%%%%%%%%%%%%%%% 
%%%%                         FIGURE 3                           %%% 
%%%%%%%%%%%%%%%%%%%%%%%%%%%%%%%%%%%%%%%%%%%%%%%%%%%%%%%%%%%%%%%%%%% 
\begin{figure} 
\psfig{figure=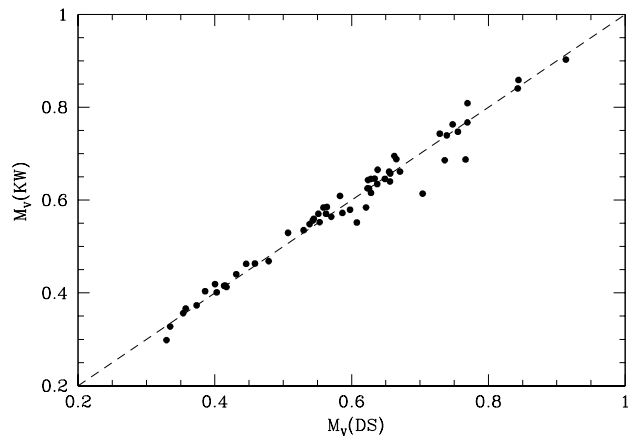,height=9cm,width=9cm} 
\caption{Comparison of the absolute 
visual magnitudes for a sample of 59 field variables (see text) obtained by using 
alternatively equation 3) and the KW relation (equation 5), once fixed its zero 
point to the same value adopted for the PLA relation.} 
\end{figure} 
%%%%%%%%%%%%%%%%%%%%%%%%%%%%%%%%%%%%%%%%%%%%%%%%%%%%%%%%%%%%%%%%%%% 
 
If one relies on the accuracy of the distance modulus
estimates obtained by using the CDP approach, then
the results obtained by using the PLA relation depend on the metallicity, since the difference between  
the CDP distance modulus and $(m-M)_V^{DS}$ is correlated with metallicity. 
Due to the fact that
neither the fundamental pulsational equation nor the adopted pulsational temperature scale 
depend on the star metal content,
it is clear that such occurrence has to be related to the assumption 
that the zero point of the PLA relation does not depend on the metallicity. 
This shows that the claims by both Castellani \& De Santis (1994)  
and DS about the fact that this zero point shows a negligible dependence  
on the heavy elements abundance, are no more supported by updated data. 
 
\section{A comparison with a $M_V -$Fourier coefficients relation.} 
 
Kov\'acs \& Walker (2001, hereinafter KW) have recently derived a set
of useful relationships 
connecting 
the absolute visual magnitude of $RR_{ab}$ stars with the Fourier parameters of the light curve. 
 
Their main result is that the fundamental contribution to $M_V$ comes
not only from  
the period P and from the first Fourier amplitude $A_1$, but there are statistically 
significant contributions also from additional higher order components, mainly from $A_3$: 
 
$$M_V= -1.880\log{P}-0.971A_1+0.909A_3+constant \eqno(5)$$ 
 
\noindent 
KW do not make any attempt to calibrate the zero point of this relation. However, due to the  
evidence that this formula fits the empirical data ($\sigma\approx0.04$ mag), they conclude 
that no important physical dependence has been missed in its derivation. 
%%%%%%%%%%%%%%%%%%%%%%%%%%%%%%%%%%%%%%%%%%%%%%%%%%%%%%%%%%%%%%%%%%% 
%%%%                         FIGURE 4                           %%% 
%%%%%%%%%%%%%%%%%%%%%%%%%%%%%%%%%%%%%%%%%%%%%%%%%%%%%%%%%%%%%%%%%%% 
\begin{figure} 
\psfig{figure=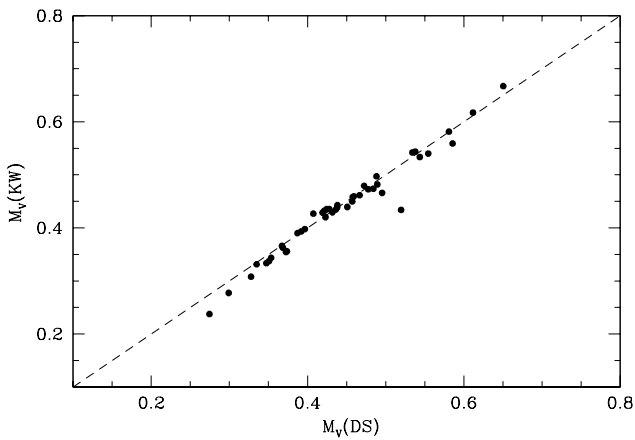,height=9cm,width=9cm} 
\caption{As figure 3, but for a sample of $RR_{ab}$  
variables in the cluster $\omega Cen$.} 
\end{figure} 
%%%%%%%%%%%%%%%%%%%%%%%%%%%%%%%%%%%%%%%%%%%%%%%%%%%%%%%%%%%%%%%%%%% 
 
We check now the consistency between the results provided  
by equation (5) and (3). We use a compilation 
of Fourier parameters for 59 field $RR_{ab}$ variables (Simon \& Teays 1982). 
In figure 3, we show the comparison between the absolute visual magnitudes obtained  by using 
equation (3) ($M_V(DS)$), and those obtained from equation (5)
($M_V(KW)$), after fixing its  
zero point to the value 0.31 mag. We find
$dM_V(KW)/dM_V(DS)=0.97\pm0.02$, with a 
correlation coefficient
r=0.98. This result clearly shows that, when the zero point of relation (5) 
is fixed consistently with the one adopted for the equation (3), the two relations are  
substantially equivalent. 
Since the KW result does not depend on any effective temperature scale 
for RR Lyrae stars, the consistency between this relation and our PLA provides 
an independent support for the reliability of the temperature-period-amplitude relation (De 
Santis 2001) adopted in the present work. It also provides an absolute calibration of the KW relation. 
 
%%%%%%%%%%%%%%%%%%%%%%%%%%%%%%%%%%%%%%%%%%%%%%%%%%%%%%%%%%%%%%%%%%% 
%%%%                         FIGURE 5                           %%% 
%%%%%%%%%%%%%%%%%%%%%%%%%%%%%%%%%%%%%%%%%%%%%%%%%%%%%%%%%%%%%%%%%%% 
\begin{figure} 
\psfig{figure=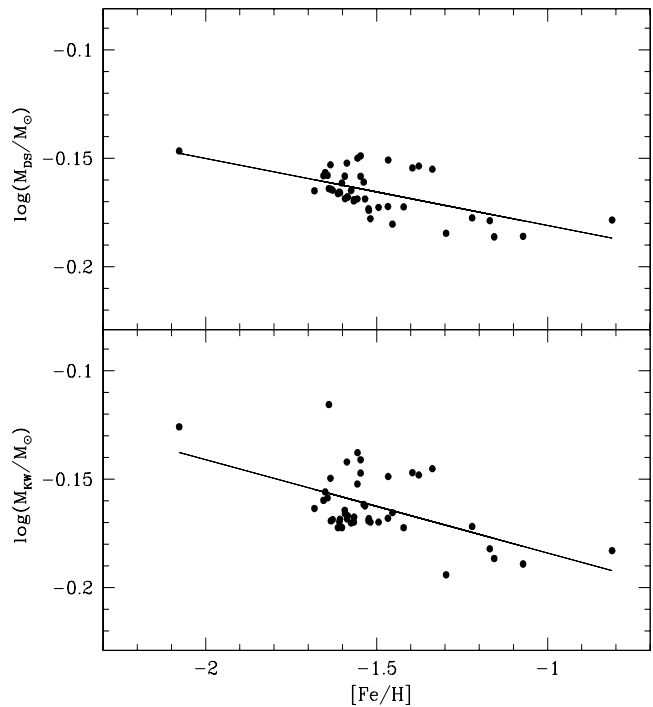,height=10cm,width=9cm} 
\caption{{\sl Panel a}: The behaviour of the logarithm of the RR Lyrae masses  
in the cluster $\omega$~Cen obtained by estimating 
the intrinsic luminosity of the variable according to equation 3, as a function 
of the metallicity. {\sl Panel b}: As in panel a), but estimating the absolute 
visual magnitude of the variables according to relation 5, once fixed its zero point 
to 0.31 mag (see text for more details).} 
\end{figure} 
%%%%%%%%%%%%%%%%%%%%%%%%%%%%%%%%%%%%%%%%%%%%%%%%%%%%%%%%%%%%%%%%%%% 
 
However, the KW relation is affected by the same metallicity 
effect as the semiempirical PLA relation. This is shown in 
Figure 4, which displays the data for 
the $RR_{ab}$ stars in the cluster $\omega$~Cen listed by KW in their table 2.  
We notice again the existence of a fine agreement between the two sets: 
$dM_V(KW)/dM_V(DS)=1.03\pm0.03$.  
In figure 5, the behaviour with the metallicity of the masses obtained by using  
either equation (3) ($M_V(DS)$) or relation (5) ($M_V(KW)$) is
shown. As for the metallicity scale, the relation  
provided by Jursik \& Kov\'acs (1996) between the metallicity and selected Fourier parameters was adopted. 
One can notice that both relations provide the same dependence 
of the mass on the metallicity, namely: $d\log{M_{KW}}/d[Fe/H]=-0.04\pm0.01$ 
and $d\log{M_{DS}}/d[Fe/H]=-0.03\pm0.006$\footnote{If the most metal-poor  
variable and the most metal-rich one are excluded from the linear fit,  
the slopes do not change significantly, both being equal to $-0.04$.}. 
 
%%%%%%%%%%%%%%%%%%%%%%%%%%%%%%%%%%%%%%%%%%%%%%%%%%%%%%%%%%%%%%%%%%% 
%%%%                         FIGURE 6                           %%% 
%%%%%%%%%%%%%%%%%%%%%%%%%%%%%%%%%%%%%%%%%%%%%%%%%%%%%%%%%%%%%%%%%%% 
\begin{figure} 
\psfig{figure=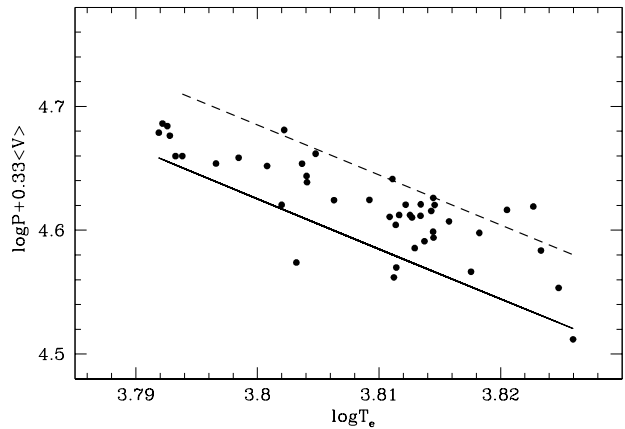,height=9cm,width=9cm} 
\caption{Comparison in the 
$(\log{P}+0.33\cdot{<V>}) - \log{T_e}$ diagram between the $RR_{ab}$ 
variables in $\omega~Cen$ and the results from 
pulsational theory, having fixed the GC distance modulus to 
$(m-M)_V=14.15$, and adopting for the allowed minimum and maximum $RR_{ab}$ mass 
the values provided by the stellar evolutionary theory. The temperature scale 
is from De Santis (1996).} 
\end{figure} 
%%%%%%%%%%%%%%%%%%%%%%%%%%%%%%%%%%%%%%%%%%%%%%%%%%%%%%%%%%%%%%%%%%% 

\section{The case of the GC $\omega$~Centauri.} 
 
It has been shown in Section 3 that the PLA relation is affected by a systematic 
metallicity effect. As a consequence, the distance  determinations are  
affected, in comparison with the CDP estimates, by a systematic uncertainty  
of the order of $-0.04$ mag for metal-poor systems and of about $+0.04$ mag for metal-rich ones. 
In the following, we show that this uncertainty is completely negligible  
when taking into account stellar systems showing a intrinsic metallicity spread. 
 
%%%%%%%%%%%%%%%%%%%%%%%%%%%%%%%%%%%%%%%%%%%%%%%%%%%%%%%%%%%%%%%%%%%  
%%%%                         FIGURE 7                           %%% 
%%%%%%%%%%%%%%%%%%%%%%%%%%%%%%%%%%%%%%%%%%%%%%%%%%%%%%%%%%%%%%%%%%% 
\begin{figure} 
\psfig{figure=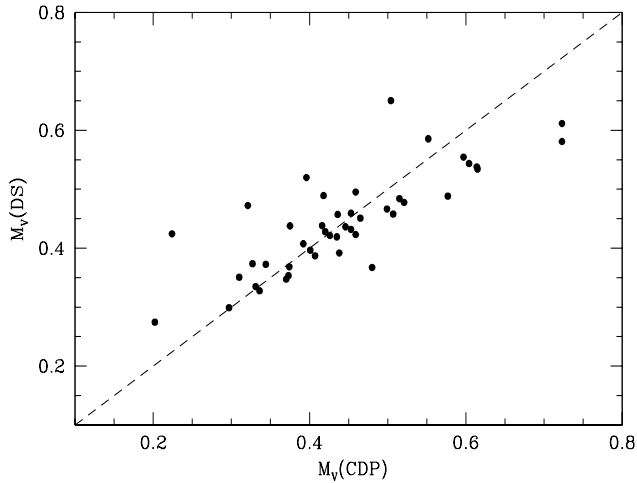,height=10cm,width=9cm} 
\caption{Comparison between the absolute 
visual magnitude obtained by using the distance modulus evaluated according 
the CDP approach (see figure 6), and the $M_V$ values obtained according the PLA 
relation (equation 3), for the RR Lyrae stars in the cluster 
$\omega~Cen$.} 
\end{figure} 
%%%%%%%%%%%%%%%%%%%%%%%%%%%%%%%%%%%%%%%%%%%%%%%%%%%%%%%%%%%%%%%%%%% 
 
We have considered the GC $\omega$~Cen, which shows an internal metallicity spread in the range $- 
2\le[Fe/H]\le-0.8$ (Rey et al. 2000 and references therein).  
According to the same evolutionary models used by DC and CDP, this  
means that the allowed mass range for $RR_{ab}$ in this cluster is  
$0.60\le{M/M_\odot}\le0.80$. By following the CDP method, we estimate a distance modulus  
$(m-M)_V=14.15\pm0.05$ mag. Figure 6 shows the behaviour  
of the reduced period as a function of the effective temperature, i.e., the empirical  
plane adopted for fixing the cluster distance modulus (see DS and CDP for details). 
By using the PLA relation for each individual RR Lyrae star, and then  
computing the mean of the obtained distance moduli, we obtain exactly the same distance  
estimate, $(m-M)_V=14.15\pm0.03$ mag. In figure 7, for each variable  
in $\omega$~Cen, the absolute visual magnitude obtained by using the CDP distance modulus  
is compared with its $M_V$ value derived from the PLA relation.  
 
%%%%%%%%%%%%%%%%%%%%%%%%%%%%%%%%%%%%%%%%%%%%%%%%%%%%%%%%%%%%%%%%%%% 
%%%%                         FIGURE 8                           %%% 
%%%%%%%%%%%%%%%%%%%%%%%%%%%%%%%%%%%%%%%%%%%%%%%%%%%%%%%%%%%%%%%%%%% 
\begin{figure} 
\psfig{figure=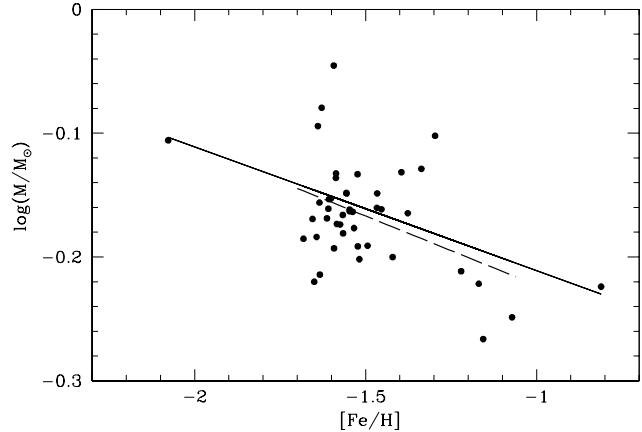,height=9cm,width=9cm} 
\caption{The behaviour of the logarithm of the RR Lyrae mass in the
cluster $\omega$~Cen  -- obtained 
by using the method discussed in section 2 and assuming a distance modulus
$(m-M)_V=14.15$ mag --  
as a function of the metallicity. 
The solid line denotes a linear regression to the whole data sample, whereas 
the dashed line represents a linear regression performed after excluding from the sample  
the most metal-poor variable and the most metal-rich one (see text for more details).} 
\end{figure} 
%%%%%%%%%%%%%%%%%%%%%%%%%%%%%%%%%%%%%%%%%%%%%%%%%%%%%%%%%%%%%%%%%%% 
 
The satisfactory agreement between the two methods  
applied to stellar systems with an internal metallicity spread, is clearly due to the  
fact that the metallicity affects the PLA relation in opposite directions for  
metal-poor and metal-rich variables. If a metallicity spread is present, the two  
effects cancel out when computing the distance modulus from the 
mean of the individual values obtained for the whole 
RR Lyrae sample. 
This occurrence makes the PLA relation very useful 
for deriving the distance to stellar systems showing an intrinsic metallicity dispersion. 
The method is simple and it does not need any preliminar estimate of the metallicity. 
 
We notice that our distance modulus estimate for the GC $\omega$~Cen is, within the  
quoted uncertainty, in good agreement with the direct distance estimate for this clusters  
obtained by Thompson et al. (2001) by using a detached eclipsing
binary  system ($(m-M)_V=14.05\pm0.11$ mag).  
 
Recently, Caputo, Degl'Innocenti \& Marconi (2001) have obtained  
a fully theoretical PLA relation based on updated pulsational models. They have applied  
this relation to $RR_{ab}$ stars in $\omega$~Cen, obtaining a distance modulus equal  
to $(m-M)_V=14.01\pm0.11$ mag. This estimate, within 
the quoted uncertainties, appears in marginal agreement with our
result. However, Caputo et 
al. (2001) have employed a different observational database for the variables in $\omega$~Cen  
in comparison with our investigation. When applying the PLA relation by Caputo et al. 
(2001) to the data provided by KW, we obtain a distance modulus estimate only $\approx0.07$ mag 
lower than the value obtained by using the semi-empirical PLA relation. 
 
In the previous section it has been shown that the RR Lyrae mass derived from the PLA absolute  
magnitudes shows a dependence on the metallicity 
$d\log{M}_{DS}/d[Fe/H]=-0.03$, in disagreement with recent stellar 
models. We address this point by considering the variables in the cluster $\omega$~Cen. 
 
By using $(m - M)_V =14.15\pm0.04$ mag, we obtain, following the usual approach, 
the mass of each variable in this cluster. In figure 8, we show the trend of the masses as a 
function of [Fe/H]. In this case we obtain $d\log{M}/d[Fe/H]= -0.10\pm
0.03$ which is, within the errors, 
in satisfactory agreement with the slope predicted by HB stellar models.  
It is worth emphasizing that an error in the adopted distance modulus does not affect  
the slope of the $\log{M} - [Fe/H]$ relation. Since there is only one star in 
the metal-poor regime ($[Fe/H]\approx-2.1$) and just one object in the metal-rich tail 
($[Fe/H]\approx-0.8$), we performed the same linear regression excluding these two 
objects: the resulting slope is $-0.11$, in agreement with that obtained  
from the whole sample. 
 
This result is important since recently, on the basis of the analysis of double-mode RR Lyrae  
in the Large Magellanic Cloud, Bragaglia et al. (2001) have found some evidences 
against the mass-metallicity relation for HB stars predicted by 
canonical stellar models. In particular, they found no significant differences in mass for RR Lyrae 
variables in GCs of different metallicity. A similar result was found also by Kov\'acs (2001) by analizing  
the properties of the double-mode RR Lyrae variables in the Sculptor dwarf galaxy.  
A more firm understanding of this issue has to wait for direct spectroscopic measurements 
of the metallicity for a larger RR Lyrae sample, but the present result does not support  
their finding. 
 
\section{Final remarks and conclusions.} 
 
By applying both the PLA relation and the CDP method to the GC $\omega$~Cen, 
we have shown that the two approaches are in agreement when applied to stellar systems  
with an internal metallicity spread. This is an advantage of the PLA relationship over 
the CDP method, since the former does not require any preliminary metallicity estimate.  
In this respect, some dwarf galaxies in the Local Group (LG) are a
fine target to exploit this technique. 
In fact, even if current metallicity measurements, based mainly on photometric indices, reveal an  
intrinsic metallicity dispersion of the order of 0.3-0.5 dex (Mateo 1998) in most of the  
LG dwarfs, more recent high-resolution spectroscopic measurements have disclosed the  
existence of larger internal spread in the heavy elements  
abundance, for instance, $\Delta[Fe/H]=0.73$ dex and 1.53 have been obtained by Shetrone,  
Cote \& Sargent (2001) for Ursa Minor and Draco, respectively.  
%%%%%%%%%%%%%%%%%%%%%%%%%%%%%%%%%%%%%%%%%%%%%%%%%%%%%%%%%%%%%%%%%%%  
%%%%                         FIGURE 9                          %%% 
%%%%%%%%%%%%%%%%%%%%%%%%%%%%%%%%%%%%%%%%%%%%%%%%%%%%%%%%%%%%%%%%%%% 
\begin{figure} 
\psfig{figure=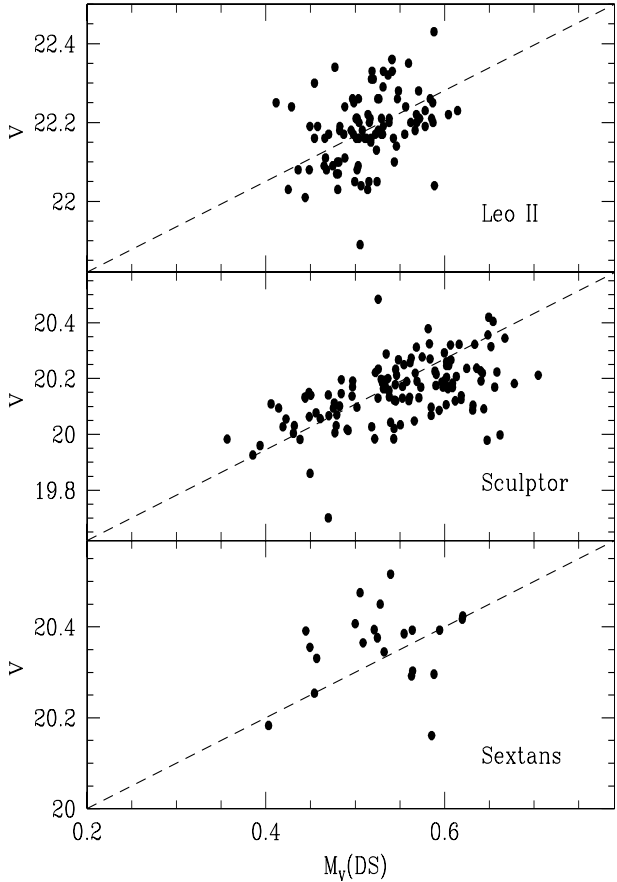,height=13cm,width=10cm} 
\caption{Comparison between the apparent visual magnitude and the  
$M_V$ value obtained from the PLA relation, for the RR Lyrae variables in the  
dwarf galaxies Leo II, Sculptor and Sextans.} 
\end{figure} 
%%%%%%%%%%%%%%%%%%%%%%%%%%%%%%%%%%%%%%%%%%%%%%%%%%%%%%%%%%%%%%%%%%% 
 
We have applied the PLA relation to the RR Lyrae sample in three LG dwarfs: Leo II, Sculptor and Sextans.  
The data are from Siegel \& Majewski (2000) for Leo II, 
from Kaluzny et al (1995) for Sculptor, and from Mateo, Fisher \& Krzeminski (1995) for Sextans. 
For each dwarf we have measured the distance modulus by applying the PLA  
relation to each individual variable and then computed the mean of the various values. 
 
The resulting distances are $21.67\pm0.05$ mag, $19.60\pm0.03$ mag, and $19.83\pm0.06$ mag  
for Leo II, Sculptor and Sextans, respectively (the listed error corresponds to the maximum  
random error, so it does not account for the observational uncertainty affecting the apparent visual 
magnitudes). These estimates are in good agreement with currently adopted values (Mateo 1998).  
In figure 9, we show the satisfactory level of agreement between the apparent visual magnitude for each 
RR Lyrae variable and the $M_V$ value provided by the PLA relation. 
 
Since the systematic metallicity effect, affecting the PLA relation, works in opposite directions, there is  
a good agreement between the distance moduli derived by using the CDP or PLA methods.  
In the case of a single metallicity  
stellar population, a reliable distance measurement should be still
achieved. In fact, the maximum systematic  
error affecting the GC distance moduli provided by the
PLA relation is of the order of $\pm0.04$ mag. 
 
However, the accuracy of the absolute magnitudes provided by the PLA relation strongly relies on the  
reliability of the adopted blue amplitude for the $RR_{ab}$ variables. This means that one has to  
exclude from the adopted data sample
those variables affected by amplitude modulation ({\sl Blazhko} effect). 
 
We wish to summarize the assumptions upon which the PLA relation zero point relies: 
 
\begin{itemize} 
\item{the validity of the adopted temperature scale. Concerning this topic,  
De Santis (2001) has shown that this relation does not significantly depend on the selected 
sample of variables, and has discussed its range of validity. 
Moreover, DC have also shown that, when this pulsational temperature scale is used, 
consistency is achieved between several sets of $T_{eff}$-color transformations  
(Buser \& Kurucz 1978, Bessel, Castelli \& Plez 1998, and Green 1988)  
and the Lub's reddening scale. However, a significant error in the zero-point of this reddening  
scale and/or of the above quoted transformations, would translate into an error for the  
zero-point of equation (3);} 
 
\item{in section 1, we have noted that the use of the van Albada \& Baker (1971) 
relation for the fundamental pulsational equation makes the PLA relation fainter by 0.05 mag 
in comparison with the Bono et al's (1997) relation;} 
 
\item{when calibrating the zero point of the PLA relation, DS adopted $0.69M_\odot$  
as the mean mass of fundamental pulsator for a metallicity $[Fe/H]=-1.5$, while this assumption  
seems to be supported by recent HB stellar models, it relies on the
reliability of the stellar model input physics.} 
\end{itemize} 
 
\noindent 
Since all the quoted assumptions seem to be reliable, we trust the accuracy 
of the semi-empirical PLA relation for deriving the distance to stellar systems such as 
galactic GCs and LG dwarf galaxies. 
 
{\bf Acknowledgments:}  
 
We warmly thank F. Caputo and M. Salaris for an accurate reading of this manuscript and for the 
suggested improvements. We also wish to thank G. Kov\'acs and A.R. Walker for their useful comments 
and suggestions on an early draft of this paper. 
Financial support for this work was provided by MURST-Cofin 2002.

\end{document}